\documentclass[preprint,aps,nofootinbib,superscriptaddress,showpacs]{revtex4}
\usepackage{epsfig}
\usepackage{bm}
\newcommand{\hszero}{h^{0s}_{\mbox{\tiny dNN}}}
\newcommand{\htzero}{h^{0t}_{\mbox{\tiny dNN}}}
\newcommand{\hone}{h^{1}_{\mbox{\tiny dNN}}}
\def\be{\begin{equation}} \def\ee{\end{equation}} \def\bea{\begin{eqnarray}}
\def\eea{\end{eqnarray}} 

\begin{document}

\title{Parity-violating polarization in $np \rightarrow d\gamma$
with a pionless effective field theory}

\author{J. W. Shin}
\affiliation{Department of Physics and Basic Atomic Energy
Research Institute, Sungkyunkwan University, Suwon 440-746, Korea}

\author{S. Ando}
\affiliation{Theoretical Physics Group, School of Physics and
Astronomy, The University of Manchester, Manchester, M13 9PL, UK}

\author{C. H. Hyun}
\email{hch@daegu.ac.kr}
\affiliation{Department of Physics Education,
Daegu University, Gyeongsan 712-714, Korea}

\date{July 23, 2009}

\begin{abstract}
We consider the two-nucleon weak interaction 
with a pionless effective field theory.
Dibaryon fields are introduced to facilitate calculations
and ensure precision in the initial and final state propagators.
Weak interactions are accounted for with the parity-violating
dibaryon-nucleon-nucleon vertices, which contain unknown 
weak dibaryon-nucleon-nucleon coupling constants.
We apply the model to the calculation of a parity-violating
observable in the neutron-proton capture at threshold.
Result is obtained up to the linear order in the unknown 
dibaryon-nucleon-nucleon coupling constants.
We compare our result to the one obtained from a hybrid calculation,
and discuss the extension to weak interactions in the few-body systems.
\end{abstract}
\maketitle

\section{Introduction}

Weak nucleon-nucleon ($NN$) interaction has recently been formulated
in the framework of effective field theory (EFT) \cite{zhu05}.
Parity-violating (PV) weak $NN$ potentials have been obtained up to 
next-to-next-to leading order (NNLO) in the pionful theory.
The weak potentials obtained from EFT have been subsequently
applied to the calculation of PV observables in the two nucleon 
systems \cite{liu07,hyun07,bertrand08},
and the results demonstrate the adequacy of perturbative scheme 
of the EFT for the description of weak $NN$ interaction.

About 40 years ago, Danilov suggested the parametrization of the
parity mixing due to the weak $NN$ interaction in terms of five PV
low energy constants \cite{danilov65, danilov71}, 
and the idea was applied to the calculation of PV observables 
in the few-nucleon systems \cite{bertrand78, bertrand80}.
In this work, we will consider PV observables in the neutron-proton
capture at threshold, where energy scale or momentum transfer
is much small compared to the pion mass.
At this small scale pion can be treated as a heavy degree of freedom.
If pions are treated as heavy degrees, we can integrate out the
pion from the theory, and then we obtain a pionless
theory where interactions are described in terms of 
only two-nucleon contact terms.
Pionless EFT for the weak $NN$ interaction
in Ref.~\cite{zhu05} is parametrized by seven 
independent PV low energy constants (LECs) at the leading order (LO),
but recently it has been shown that two terms are redundant and thus
five terms are independent in practice \cite{girl08}.
Thus, inasmuch as the number of unknown weak parameters is concerned,
Danilov's idea in the past gives the same parametrization
to the pionless EFT at leading order.

Parity-violating vertex in the pionless theory in \cite{zhu05}
consists of the multiplication of two two-nucleon fields:
one in $S$ state and the other in $P$ state.
Given a rule to transform a two-nucleon state to the corresponding
dibaryon field, it is straightforward to obtain the PV Lagrangian
that describes the weak $NN$ interaction in terms of PV 
dibaryon-nucleon-nucleon ($dNN$) or dibaryon-dibaryon vertices.
Introducing a dibaryon field for the deuteron, 
the effective range contribution ($\gamma \rho_d \sim 0.4$) 
to the deuteron propagator is taken into account up to infinite order,
and it consequently makes the convergence of the theory improved
compared to the pionless EFT that does not have dibaryon fields.
Since scattering lengths and effective ranges in the $S$ state are
unusually large, resummation of effective range contribution to 
infinite order in dibaryon formalism is especially useful for the 
two-nucleon systems dominated by $S$ state.
In this work, we obtain the PV Lagrangian with dibaryon fields
by transforming the two-nucleon $S$ states to the corresponding
dibaryon fields, while describing the $P$ states in terms of the
two-nucleon fields.
Weak $NN$ interaction is described by the PV $dNN$ vertices,
which have unknown weak coupling constants.

We plug the Lagrangians in the calculation of the PV polarization 
($P_\gamma$) in $np \rightarrow d\gamma$ at threshold. 
PV polarization has been calculated with the weak one-meson-exchange (OME)
potentials (conventionally referred to as DDH potential \cite{ddh80})
and with various strong interaction models \cite{hyun05}.
The results in \cite{hyun05} show strong dependence on the strong
interaction model, and are dominated by the $\rho$- and $\omega$-meson
exchange terms in the DDH potential.
In the EFT, $\rho$, $\omega$ and heavier mesons are integrated out 
because their masses are very large scales at low-energy few-body 
processes, and their contributions are embedded in the $NN$ contact terms.
Since the PV polarization in $np \rightarrow d\gamma$ is dominated
by the heavy mesons in the OME picture for the weak $NN$ interaction,
if it is considered in the EFT, only the contact terms are relevant 
and thus the pionless EFT may be one of the most favorable frameworks for 
the investigation.
Result for $P_\gamma$ is obtained in terms of the unknown
weak $dNN$ coupling constants, which have to be determined
from the measurements for the relevant PV observables.

We outline the paper as follows. 
In Sec.~II, we present the parity-conserving and 
the parity-violating Lagrangians that contribute to the
observable at leading order. 
In Sec.~III, we obtain the PV polarization in 
unpolarized neutron capture by a proton at threshold,
and discuss the result. We conclude the paper in Sec.~IV.

\section{Effective Lagrangian}
Parity-conserving (PC) Lagrangian includes strong
and electromagnetic (EM) interactions.
PC Lagrangian with dibaryon fields can be written as 
\begin{eqnarray}
{\cal L}_{\rm PC} = {\cal L}_N + {\cal L}_s + {\cal L}_t 
+ {\cal L}_{st},
\end{eqnarray}
where ${\cal L}_N$, ${\cal L}_s$, ${\cal L}_t$ and ${\cal L}_{st}$ 
represent PC interactions for nucleons, dibaryon in $^1 S_0$ state, 
dibaryon in $^3 S_1$ state, and EM transition between 
$^1 S_0$ and $^3 S_1$ states, respectively. 
Retaining the terms that are relevant to the quantity of
interest in this work, we have
\begin{eqnarray}
{\cal L}_N &=& N^\dagger \left( i v \cdot D + \frac{1}{2 m_N}
\left\{ (v \cdot D)^2 - D^2 \right\}  \right) N, \\
{\cal L}_s &=& \sigma_s s^\dagger_a \left\{ i v \cdot D + \frac{1}{4 m_N}
\left[ ( v \cdot D)^2 - D^2 \right] + \Delta_s \right\} s_a
- y_s \left\{ s^\dagger_a [ N^T P^{(^1 S_0)}_a N ] + {\rm h.c.} \right\}, \\
{\cal L}_t &=& \sigma_t t^\dagger_i \left\{ i v \cdot D + \frac{1}{4 m_N}
\left[ ( v \cdot D)^2 - D^2 \right] + \Delta_t \right\} t_i
- y_t \left\{ t^\dagger_i [ N^T P^{(^3 S_1)}_i N ] + {\rm h.c.} \right\}, \\
{\cal L}_{st} &=& \frac{L_1}{m_N \sqrt{r_0 \rho_d}} 
[ t^\dagger_i s_3 B_i + {\rm h.c.} ],
\end{eqnarray}
where the projection operators for the $^1 S_0$ and $^3 S_1$ states
are defined respectively as
\begin{eqnarray}
P^{(^1 S_0)}_a &=& \frac{1}{\sqrt{8}} \sigma_2 \tau_2 \tau_a, \\
P^{(^3 S_1)}_i &=& \frac{1}{\sqrt{8}} \sigma_2 \sigma_i \tau_2.
\end{eqnarray}
Velocity vector $v_\mu$ satisfies $v^2 = 1$, and
$D_\mu =\partial_\mu - i{\cal V}_\mu^{\rm ext}$ where
${\cal V}^{\rm ext}_\mu$ represents the external vector field.
Dibaryon fields in $^1 S_0$ and $^3 S_1$ states are denoted
by $s_a$ and $t_i$, respectively, and $B_i$ is external magnetic
field given by $\vec{B} = \nabla \times \vec{{\cal V}}^{\rm ext}$.
$\sigma_s$ and $\sigma_t$ are the sign factors having a value $-1$,
and $\Delta_{s,t}$ are defined by the mass difference between the
dibaryon and two nucleon states as $\Delta_{s, t} = m_{s, t} - 2 m_N$. 
Low energy constants $y_s$ and $y_t$ are the strong $dNN$ 
coupling constants 
determined from the empirical values of effective ranges.
We obtain $y_s = \frac{2}{m_N} \sqrt{\frac{2 \pi}{r_0}}$ and
$y_t = \frac{2}{m_N} \sqrt{\frac{2 \pi}{\rho_d}}$, where
$r_0$ is the effective range in $^1 S_0$ state and $\rho_d$ is
the effective range for the deuteron. 
LEC $L_1$ denotes the photon-dibaryon-dibaryon coupling constants
for the M1 transition, and it has to be determined from experiments.

PV Lagrangian for the two nucleon system
can be written as
\begin{eqnarray}
{\cal L}_{\mbox{\tiny PV}} = \sum_{\Delta I} 
{\cal L}^{\Delta I}_{\mbox{\tiny PV}}
\end{eqnarray}
where $\Delta I$ denotes the isospin change in the PV vertex.
PV vertex changes the orbital angular momentum by
an odd number (e.g, $S \leftrightarrow P$).
Because $\Delta (L + S + I)$ has to be even,
we have $\Delta (S + I) = 1$ for the two nucleon system.
Consequently we have
\begin{eqnarray}
{\cal L}_{\mbox{\tiny PV}} = 
{\cal L}^0_{\mbox{\tiny PV}} + {\cal L}^1_{\mbox{\tiny PV}}.
\end{eqnarray}
Since the total angular momentum is conserved in the $NN$ interaction, 
parity mixings allowed by the PV interaction
for the lowest orbital states are
$^1 S_0 \leftrightarrow {}^3 P_0$, and $^3 S_1 \leftrightarrow {}^1 P_1$
due to ${\cal L}^0_{\mbox{\tiny PV}}$, and $^3 S_1 \leftrightarrow {}^3 P_1$
due to ${\cal L}^1_{\mbox{\tiny PV}}$.
In the pionless theory interaction is described only by
the nucleon-nucleon contact terms which have undetermined LECs.
In the case of pionless theory with dibaryon fields, we assume
that a PV $dNN$ vertex subsumes the PV $NN$ interactions.
Non-relativistic P-odd and T-even Lagrangian for the neutron-proton
system with $\Delta I = 0$ can be written as
\begin{eqnarray}
{\cal L}^0_{\mbox{\tiny PV}} &=&
\frac{h^{0 s}_{\mbox{\tiny dNN}}}{2 \sqrt{2\, \rho_d\, r_0}\, m_N^{5/2}} 
s^\dagger_3\, N^T 
\sigma_2 \sigma_i  \tau_2 \tau_3 
\frac{i}{2} \left(\stackrel{\leftarrow}\nabla - 
\stackrel{\rightarrow}\nabla \right)_i N +{\rm h.c.} \label{eq:Lwk0s}\\
& & + 
\frac{h^{0 t}_{\mbox{\tiny dNN}}}{2 \sqrt{2} \rho_d\, m_N^{5/2}} \,
t^\dagger_i\, N^T \sigma_2 \tau_2 
\frac{i}{2} \left(\stackrel{\leftarrow}\nabla - 
\stackrel{\rightarrow}\nabla \right)_i N
+{\rm h.c.}, \label{eq:Lwk0t}
\end{eqnarray}
where $\hszero$ and $\htzero$ denote the weak $dNN$ coupling
constants for the parity mixing for the $^1 S_0$ and $^3 S_1$
states, respectively.
Spin and isospin operators $\sigma_2 \sigma_i \tau_2 \tau_a$ in 
Eq.~(\ref{eq:Lwk0s}) projects two-nucleon system to $^3 P_0$
state. PV vertex given by Eq.~(\ref{eq:Lwk0s}) therefore generates
$^3 P_0$ admixture in the $^1 S_0$ state. 
Similarly, $\sigma_2 \tau_2$ in Eq.~(\ref{eq:Lwk0t}) is the
projection operator for $^1 P_1$ state, and thus the Lagrangian
mixes $^1 P_1$ state in the $^3 S_1$ state.
For the $\Delta I= 1$ part, we have $^3 P_1$ admixture to the $^3 S_1$
state, so the Lagrangian reads
\begin{eqnarray}
{\cal L}^1_{\mbox{\tiny PV}} =
i \frac{h^1_{\mbox{\tiny dNN}}}{2 \sqrt{2} \rho_d\, m_N^{5/2}} \,
\epsilon_{ijk}\, t^\dagger_i\, N^T \sigma_2 \sigma_j \tau_2  \tau_3
\frac{i}{2} \left(\stackrel{\leftarrow}\nabla - 
\stackrel{\rightarrow}\nabla \right)_k N
+{\rm h.c.}. \label{eq:Lwe1t}
\end{eqnarray}

Lagrangians given in Eqs.~(\ref{eq:Lwk0s},\ref{eq:Lwk0t},\ref{eq:Lwe1t}) 
represent weak interactions between a neutron and a proton.
Full LO interactions in the pionless theory, which include 
$nn$ and $pp$ weak interactions as well as the $np$ one
can be found in the literature \cite{dan08}.
By transforming a two-nucleon field in $S$ state
to the corresponding dibaryon field,
one can easily get a mapping between pionless theories with and
without the dibaryon fields.
We will discuss the relation of the two theories
in more detail in the discussion of the result.


\section{Result and Discussion}

In the pionless theory, expansion parameters are
$Q/m_\pi$ or $Q/\Lambda$, where $Q$ is a small momentum,
$m_\pi$ the pion mass and $\Lambda$ a symmetry breaking scale.
Since the scattering lengths and effective ranges in the $^1 S_0$
and $^3 S_1$ states are large, we count their inverse as small scales,
i.e. $(\gamma,\,\, 1/a_s,\,\, 1/a_t,\,\, 1/r_0,\,\, 1/\rho_d) \sim Q$, 
where $a_{s(t)}$ is the scattering length in $^1 S_0 (^3 S_1)$ state 
and $\gamma = \sqrt{m_N B}$ with $B$ the deuteron binding energy.
Nucleon and dibaryon propagators are counted as $1/Q^2$ 
and a loop integral contributes an order of $Q^5$.

\begin{figure}[tbp]
\epsfig{file=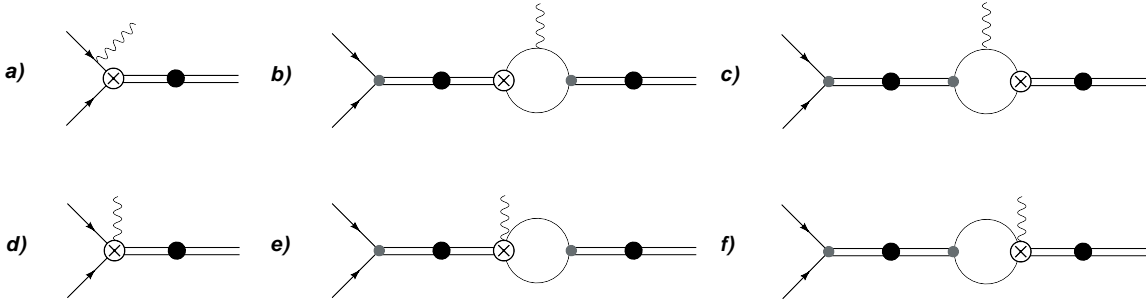, width=6in}
\caption{Leading order ($Q^0$) PV diagrams for $np$ capture.
Single solid line denotes a nucleon, wavy line a photon, and
a double line with a filled circle stands for dressed dibaryon propagator.
Circle with a cross represents a PV $dNN$ vertex.} 
\label{fig:pvnpdg}
\end{figure}

Feynman diagrams at leading order are depicted in Fig.~\ref{fig:pvnpdg},
which are of the order of $Q^0$.
Single solid and wavy lines represent nucleon and photon fields, respectively.
Double line with filled circle denotes the dressed dibaryon fields,
which includes the infinite sum of the intermediate nucleon loops.
Small dot at the dibaryon-nucleon-nucleon vertex denotes the 
strong $dNN$ coupling, which is proportional to $y_s$ or $y_t$, 
and the circle with a cross denotes the PV $dNN$ vertex proportional to 
$\hszero$, $\htzero$ and $\hone$.  
For the photon-nucleon coupling in Fig.~\ref{fig:pvnpdg} (a-c),
we employ the vertex function of the convection current given by
\begin{eqnarray}
i \Gamma_{VNN}(E1) =  \frac{i}{2 m_N} (1 + \tau_3) 
\frac{1}{2} (\vec{p}+\vec{p}') \cdot \vec{\epsilon}^*_\gamma,
\label{eq:E1ham}
\end{eqnarray}
where $\vec{p}$ and $\vec{p}'$ are the in-coming and out-going
nucleon momentum at the photon-nucleon vertex, respectively,
and $\vec{\epsilon}^*_\gamma$ is the polarization of out-going
photons.
For the PV photon-dibaryon-nucleon-nucleon ($VdNN$) vertex 
in Fig.~\ref{fig:pvnpdg} (d-f),
we assume minimal coupling to the PV $dNN$ vertex,
\begin{eqnarray}
\vec{\nabla} \to \vec{\nabla} - i \frac{e}{2} (1 + \tau_3) \vec{V},
\end{eqnarray}
where $\vec{V}$ denotes the external photon field.
With the minimal coupling, coupling constants at the $VdNN$ 
vertices are the same with those at the $dNN$ ones.
Resulting amplitudes are, therefore, proportional to the weak 
$dNN$ coupling constants $\hszero$, $\htzero$ or $\hone$, and
thus we have three unknown coefficients in the result.

PV polarization $P_\gamma$ in $np \to d\gamma$ is defined as
\begin{eqnarray}
P_\gamma = \frac{\sigma_+ - \sigma_-}{\sigma_+ + \sigma_-},
\label{eq:polarization}
\end{eqnarray}
where $\sigma_+$ and $\sigma_-$ are the total cross section for
the photons with right and left helicity, respectively.
$P_\gamma$ was measured in 70's, and the reported value is 
$P_\gamma = (1.8 \pm 1.8) \times 10^{-7}$ \cite{pgexp84},
but there was no more measurement after that.
At threshold, PV asymmetry in $d \vec{\gamma} \to np$ is
equal to PV polarization in $np \to d\gamma$.
PV asymmetry in $d\vec{\gamma} \rightarrow np$ has been recently
calculated with the DDH potential up to about 10 MeV above threshold
\cite{liu04,fuji04}.
Absolute value of the asymmetry is maximum at threshold
and it decreases very quickly as the energy increases.
Measurement may be most feasible at threshold, and 
if the measurement is performed at threshold,
it can be directly related to the PV polarization in
$np \rightarrow d\gamma$.

Transition amplitude that includes both PC and PV contributions
can be written as
\begin{equation}
i M_{np} = \left[ 
Y \vec{\epsilon}^*_d \cdot (\hat{k} \times \vec{\epsilon}^*_\gamma)
- i Z \vec{\epsilon}^*_d \cdot \vec{\epsilon}^*_\gamma \right]
N^T P^{(^1 S_0)}_3 N. 
\end{equation}
$Y$ denotes the PC amplitude, and we take the result in Ref.~\cite{ando05},
\begin{eqnarray}
Y &=& \frac{\sqrt{2 \pi}}{m^2_N} \sqrt{\frac{\gamma}{1 -\gamma \rho_d}}
\left[ (1 + \kappa_V) ( 1 - \gamma a_s) - \gamma^2 a_s L_1 \right],
\label{eq:pcamp}
\end{eqnarray}
where $\kappa_V$ (= 3.706) is the isovector anomalous magnetic moment
of the nucleon, $\gamma$ ($= \sqrt{m_N B} = 45.7$ MeV) is the deuteron
momentum, $\rho_d$ ($= 1.764$ fm) is the deuteron effective range,
and $a_s$ ($= -23.732$ fm) is the neutron-proton scattering length
in the $^1 S_0$ state.
We can reproduce the neutron-proton capture cross section at threshold, 
$\sigma_{\rm exp} = 334.2 \pm 0.5$ mb 
with $L_1 = - 4.427 \pm 0.015$ fm \cite{ando05}.
$Z$ is the PV amplitude for the transition from initial $^1 S_0$
to final $^3 S_1$ states. 
PV polarization $P_\gamma$ is obtained in terms of PC and PV amplitudes as
\begin{equation}
P_\gamma = -2 \frac{{\rm Re} (Y Z^*)}{|Y|^2}.
\end{equation}
We obtain the PV amplitudes for the diagrams in Fig.~\ref{fig:pvnpdg} as
\begin{eqnarray}
Z_a &=& - \frac{1}{3}\frac{\htzero}{m^2_N \sqrt{m_N \rho_d}}
\sqrt{\frac{\gamma}{1-\gamma\rho_d}}
\frac{p^2}{\gamma^2+p^2}, \\
Z_b &=& - \frac{1}{3}\frac{\hszero}{m^2_N \sqrt{m_N \rho_d}}
\sqrt{\frac{\gamma}{1-\gamma\rho_d}} 
\frac{1}{\frac{1}{a_s}-\frac{1}{2}r_0 p^2+ip}\frac{\gamma^3+ip^3}{\gamma^2+p^2}, \\
Z_c &=& - \frac{1}{3}\frac{\htzero}{m^2_N \sqrt{m_N \rho_d}}
\sqrt{\frac{\gamma}{1-\gamma\rho_d}} 
\frac{1}{\frac{1}{a_s}-\frac{1}{2}r_0 p^2+ip}\frac{\gamma^3+ip^3}{\gamma^2+p^2}, \\
Z_d &=& \frac{1}{2} \frac{\htzero}{m^2_N \sqrt{m_N \rho_d}}
\sqrt{\frac{\gamma}{1-\gamma\rho_d}}, \\
Z_e &=& \frac{1}{2} \frac{\hszero}{m^2_N \sqrt{m_N \rho_d}}
\sqrt{\frac{\gamma}{1-\gamma\rho_d}} 
\frac{\gamma}{\frac{1}{a_s} - \frac{1}{2} r_0 p^2 + ip},\\
Z_f &=& -\frac{1}{2} \frac{\htzero}{m^2_N \sqrt{m_N \rho_d}}
\sqrt{\frac{\gamma}{1-\gamma\rho_d}}
\frac{ip}{\frac{1}{a_s} - \frac{1}{2} r_0 p^2 + ip},
\end{eqnarray}
where $r_0$ ($=2.70$ fm) is the neutron-proton effective range 
in the $^1 S_0$ channel.
Taking the limit $p \to 0$ at threshold, we obtain the net PV amplitude $Z$, 
\begin{eqnarray}
Z = \frac{1}{m^2_N \sqrt{m_N \rho_d}} 
\sqrt{\frac{\gamma}{1-\gamma\rho_d}} 
\left[\htzero \left( \frac{1}{2} - \frac{1}{3} \gamma a_s \right)
+ \frac{1}{6} \hszero \gamma a_s \right],
\label{eq:pvz}
\end{eqnarray}
and the PV polarization $P_\gamma$ at LO reads
\begin{eqnarray}
P_\gamma &=& - \sqrt{\frac{2}{\pi m_N \rho_d}} 
\frac{\left(\frac{1}{2}-\frac{1}{3}\gamma a_s\right) \htzero 
+ \frac{1}{6} \gamma a_s \hszero }
{(1+\kappa_V)(1-\gamma a_s)-\gamma^2 a_s L_1} \nonumber \\
&=& -(2.59 \htzero - 1.01 \hszero) \times 10^{-2}.
\label{eq:pgamma}
\end{eqnarray}
PV polarization turns out to depend on two weak coupling constants 
$\htzero$ and $\hszero$, 
and thus we cannot determine them uniquely from a single measurement 
of $P_\gamma$ at threshold. 
In order to determine them unambiguously, 
we need more data for $P_\gamma$ at energies other than threshold, 
or measurements of observables that are independent of $P_\gamma$. 
We will discuss this matter in more detail in the conclusion.

Now we try to compare our result to the one 
obtained with a pionless theory where there is no dibaryon field \cite{liu07}.
We start from the pionless PV Lagrangian in Ref.~\cite{dan08}.
If we transform two-nucleon fields in $S$ state in Eq.~(6) in Ref.~\cite{dan08} 
to a dibaryon field, we obtain the PV Lagrangian in the dEFT given
by Eqs.~(\ref{eq:Lwk0s}-\ref{eq:Lwe1t}).
We use the transformations from two-nucleon fields to a single dibaryon 
one given by
\begin{eqnarray}
N^T P^{(^1 S_0)}_a N \to \frac{y_s}{C^{(^1 S_0)}_0} s_a ,\,\,\,
N^T P^{(^3 S_1)}_i N \to \frac{y_t}{C^{(^3 S_1)}_0} t_i, 
\label{eq:trans}
\end{eqnarray}
for the $^1 S_0$ and $^3 S_1$ states, respectively.
$C^{(^1 S_0)}_0$ and $C^{(^3 S_1)}_0$ are the coefficients for the LO
strong two-nucleon contact terms in the pionless theory.
In the power divergence subtraction scheme, they are given as
\begin{eqnarray}
\frac{1}{C^{(^1 S_0)}_0} = \frac{m_{N}}{4\pi}\left(\frac{1}{a_0} - \mu\right), 
\,\,\,\,
\frac{1}{C^{(^3 S_1)}_0} = \frac{m_{N}}{4\pi}\left(\gamma - 
\frac{1}{2} \gamma^2 \rho_d -\mu\right),
\end{eqnarray}
where $\mu$ is the renormalization point.
Substituting the transformations given by Eq.~(\ref{eq:trans}) into
Eq.~(6) in Ref.~\cite{dan08}, and comparing them with the PV dEFT Lagrangians
in Eqs.~(\ref{eq:Lwk0s}, \ref{eq:Lwk0t}), we obtain 
\begin{eqnarray}
\hszero &=& 16 \sqrt{\frac{\rho_d m_N}{2 \pi}} m^2_N
\left( \frac{1}{a_0} - \mu\right) \left(
{\cal C}^{(^1S_0- ^3P_0)}_{\Delta I =0}
- 2 {\cal C}^{(^1S_0- ^3P_0)}_{\Delta I =2} \right), \label{eq:trans0s} \\
\htzero &=& 16 \sqrt{\frac{\rho_d m_N}{2 \pi}} m^2_N
\left( \gamma - \frac{1}{2} \gamma^2 \rho_d - \mu\right) 
{\cal C}^{(^3S_1- ^1P_1)}. 
\label{eq:trans0t}
\end{eqnarray}
Inserting Eqs.~(\ref{eq:trans0s}, \ref{eq:trans0t}) to
the result for PV amplitude in Eq.~(\ref{eq:pvz}) and
assuming $\mu = m_\pi$, we obtain
\begin{eqnarray}
Z \propto  {\cal C}^{(^3S_1- ^1P_1)}  
- 0.56 \left({\cal C}^{(^1S_0- ^3P_0)}_{\Delta I =0}
- 2 {\cal C}^{(^1S_0- ^3P_0)}_{\Delta I =2}\right).
\label{eq:pvzlec}
\end{eqnarray}
Isospin change is zero at the vertex denoted by 
${\cal C}^{(^3 S_1 - {}^1 P_1)}$, i.e. $\Delta I =0$, 
and thus assuming roughly ${\cal C}^{(^3S_1- ^1P_1)} \sim
{\cal C}^{(^1S_0- ^3P_0)}_{\Delta I =0}$, we obtain the
ratio of the coefficient for $\Delta I = 0$ contribution
to that for $\Delta I = 2$ one in Eq.~(\ref{eq:pvzlec})
approximately one half.
PV polarization has been calculated in the hybrid scheme in Ref.~\cite{liu07}, 
where strong interaction is described by Argonne $v18$ model (A$v18$), 
weak interaction by the pionless EFT and the EM operator by
Siegert theorem. 
The result in Ref.~\cite{liu07} is represented in terms of Danilov parameters.
Substituting the relations of Danilov parameters and PV LECs
in the pionless theory to the result, $P_\gamma$ reads
\begin{eqnarray}
P_\gamma(\mbox{hybrid}) = (-0.25 C_1 + 2.14 C_3 + 4.18 C_5) \times 10^{-3},
\label{eq:pgliu}
\end{eqnarray}
where $C_1$ and $C_3$ correspond to $\Delta I = 0$ vertices and 
$C_5$ to $\Delta I = 2$ one. 
Assuming $C_3 \sim C_1$ and comparing the coefficients 
for $\Delta I =0$ contribution
to that of $\Delta I = 2$ in Eq.~(\ref{eq:pgliu}),
we obtain a ratio roughly one half, which is similar to our result.
Similar value of the ratio
has also been obtained from the calculation with DDH potential 
for the weak interaction and A$v18$ for the strong one \cite{hyun05}.

\section{Conclusion}

We have calculated the PV  polarization in $np \rightarrow d\gamma$ 
at the threshold with a pionless EFT with dibaryon fields.
Weak $NN$ interactions are described with the PV $dNN$ vertices, 
and the PV observable has been obtained in terms of the 
PV $dNN$ coupling constants. Precise measurement of the observable
will provide a constraint to 
determine the PV coupling constants unambiguously.

EFT has been employed partially in the calculation of PV asymmetry
in $\vec{n} p \rightarrow d\gamma$ with the pionful theory 
\cite{liu07, hyun07, hyun01, schi03}.
For instance, in Ref.~\cite{hyun01, schi03}, meson-exchange currents (MECs) 
are obtained up to an order, but the strong interaction is
described by a phenomenological model, A$v18$, and weak interaction by
the DDH potential.
In Refs.~\cite{liu07, hyun07}, on the other hand, weak potential is 
expanded up to a given order, while EM operator is accounted for 
with Siegert theorem and strong interaction described with A$v18$.
Calculation where EFT is partially employed is called hybrid calculation.
Current conservation for a given PV potential and the corresponding
MEC has been used as a crucial criterion in the calculation of
the anapole moment of the deuteron \cite{hyun03, liu03}.
Since the current conservation can be satisfied when a potential and
corresponding MECs are taken into account consistently, 
consistent expansion of strong potentials, weak potentials and
transition operators is an important requirement in the EFT.
It has been pointed out that the orders of the interactions and
transition operators in the hybrid calculations are in serious
disagreement \cite{dan08}.
In our calculation with pionless dEFT, the order of a diagram
is obtained by counting the strong, weak and EM vertices altogether,
and we truncate the expansion at a given order.
Therefore, our calculation satisfies the consistency requirement
mentioned above.
On the other hand, results from the conventional calculation, where
strong interaction is accounted with modern potential, weak interaction
with DDH potential, and EM operator with Siegert theorem, 
have provided benchmarks to both experiment and theory,
but the physical criteria such as current conservation have seldom
been checked carefully.
It is important to understand the uncertainty due to
the order mismatch in the conventional and hybrid calculations, 
and investigation along this direction with either pionful or pionless 
theories is an important future work.

There are five weak LECs in the pionless theory and therefore
we need at least five data for the PV observables.
$P_\gamma$ may be one of them.
Recently PV longitudinal asymmetries in $\vec{p}p$, $\vec{n} p$ and
$\vec{n} n$ scattering have been calculated with a pionless EFT \cite{dan08}.
Longitudinal asymmetry in $\vec{p}p$ depends on three PV coupling constants
${\cal C}^{(^1 S_0 -{}^3 P_0)}_{\Delta I = 0, 1, 2}$, and thus the measurement
at 13.6 MeV provides a relation for them.
PV asymmetry in $\vec{n} p \rightarrow d\gamma$ and deuteron anapole moment
have been calculated with the pionless dEFT \cite{savage01}, and the 
results turn out to be dominated by $\hone$.
Measurement of the PV asymmetry at SNS is expected to play an important
role in determining the value of weak LEC $\hone$ (or the weak
pion-nucleon coupling constant $h^1_\pi$).
Turning to the possibilities in the three body system, one can find a
recent calculation of the weak effect in the spin rotation in $\vec{n} d$
scattering \cite{schi08}. The authors employed DDH potential for the
weak interaction, and obtained a result dominated by $h^1_\pi$.
Though the asymmetry in $\vec{n}p$ and spin rotation in $\vec{n} d$ are
observables independent to each other, they are exclusively dependent
on $h^1_\pi$ (or equivalently $\hone$), so the measurements of the 
observables will provide a check for the consistency of $h^1_\pi$.
In order to determine the remaining weak LECs in the pionless EFT,
we need calculations and measurements for as many observables as possible.

Among many possible PV observables in the few-body systems, 
an observable that draws our interest is the PV asymmetry in 
$\vec{n} d \to t\gamma$ 
at threshold.
It has been measured at ILL \cite{avenir84},
and the reported result reads
\[
A^t_\gamma = (4.2 \pm 3.8) \times 10^{-6}.
\]
Theoretical calculation of $A^t_\gamma$ in Ref.~\cite{bertrand86} adopted
DDH potential for the weak interaction, and examined the dependence
on the strong interaction models such as de Tourreil-Sprung (TS) and
Reid soft core (RSC).
The results are interesting in some aspects. 
First, dependence on the strong interaction model is non-negligible;
TS model gives a result
$A^t_\gamma(\mbox{TS}) = 0.81 \times 10^{-6}$ 
while RSC gives $A^t_\gamma(\mbox{RSC}) = 0.61 \times 10^{-6}$.
Second, isoscalar, isovector and isotensor PV interactions in the
DDH potential give similar contributions to $A_\gamma^t$, e.g.
$0.40$, $0.45$, and $-0.04$, respectively, with the TS model. 
This means that contributions from $\pi$, $\rho$ and $\omega$
exchanges in the PV potential are similar to each other. 
Dependence on the strong model and the non-negligible contribution
from the heavy mesons are the features common with the 
PV polarization in $np \rightarrow d\gamma$.
In this problem again, therefore, pionless EFT will serve us with a most
natural and systematic way to parametrize the parity-mixing in the 
few-body system due to weak nuclear force.

Parity violation in the three- and few-body systems can show us
the effects that are not accessible in the two-body systems.
For instance, strong $3N$ force can give non-negligible correction
to the one- and two-body contributions to the PV observables. 
There has been no consideration on the weak three-body force, 
but we have recently obtained non-zero component of weak $3N$ force 
in a preliminary calculation \cite{song09}.
Two- and three-body PV meson-exchange currents are also important issues.
We expect that the EFT will play a crucial role in extending our 
understanding of the nuclear weak force in few-body systems.

%
%

\section*{Acknowledgments}

We are grateful to B. Desplanques for reading the manuscript and
comments on it.
This research was supported by the Daegu University Research Grant, 2008.

\end{document}